\documentclass[aps,prl,twocolumn,10pt,superscriptaddress,showpacs]{revtex4-1}

\usepackage{graphicx}
\usepackage{hyperref}
\usepackage{url}
\usepackage{amsmath}
\usepackage{verbatim}
\usepackage{color}

\def\beq {\begin{equation}}
\def\eeq {\end{equation}}
\def\w {\omega}
\def\bfr {\mathbf{r}}
\def\bfk {\mathbf{k}}
\def\bfq {\mathbf{q}}

\def\bfG {\mathbf{G}}

\newcommand{\bra}[1]{\langle #1|}
\newcommand{\ket}[1]{|#1\rangle}

\date{\today}

\begin{document}
\title{
Exciton energy-momentum map of hexagonal boron nitride}

\newcommand{\lsi}{Laboratoire des Solides Irradi\'es, \'Ecole Polytechnique, CNRS, CEA-DSM-IRAMIS,  Universit\'e Paris-Saclay, F-91128 Palaiseau, France}
\newcommand{\etsf}{European Theoretical Spectroscopy Facility (ETSF)}
\newcommand{\soleil}{Synchrotron SOLEIL, L'Orme des Merisiers, Saint-Aubin, BP 48, F-91192 Gif-sur-Yvette, France}
\newcommand{\helsinki}{Department of Physics, P.O. Box 64, FI-00014 University of Helsinki, Helsinki, Finland}
\newcommand{\japan}{National Institute for Materials Science, Namiki 1-1, Tsukuba, Ibaraki 305-0044, Japan}

\author{Giorgia Fugallo}
\affiliation{\lsi}
\affiliation{\etsf}
\author{Matteo Aramini} 
\affiliation{\helsinki}
\author{Jaakko Koskelo}
\affiliation{\helsinki}
\author{Kenji Watanabe} 
\affiliation{\japan}
\author{Takashi Taniguchi} 
\affiliation{\japan}
\author{Mikko Hakala}
\affiliation{\helsinki}
\author{Simo Huotari}
\affiliation{\helsinki}
\author{Matteo Gatti}
\affiliation{\lsi}
\affiliation{\etsf}
\affiliation{\soleil}

\author{Francesco Sottile}
\affiliation{\lsi}
\affiliation{\etsf}

\begin{abstract}

Understanding and controlling the way excitons propagate in solids is a key for tailoring materials with improved optoelectronic properties.
A fundamental step in this direction is the determination of the exciton energy-momentum dispersion.
Here, thanks to the solution of the parameter-free Bethe-Salpeter equation (BSE),
we draw and explain the exciton energy-momentum map  of hexagonal boron nitride (h-BN) in the first three Brillouin zones.
We show that h-BN displays strong excitonic effects 
 not only in the optical spectra at vanishing momentum  $\bfq$, 
as previously reported, but also  at large $\bfq$. 
We validate our theoretical predictions  by assessing the calculated exciton map by means of an inelastic x-ray scattering (IXS) experiment.
Moreover, we solve the discrepancies between previous experimental data and calculations, proving then that the BSE is highly accurate
through the whole momentum range.
Therefore, these results put forward the combination BSE and IXS as the tool of choice for addressing the exciton dynamics in complex materials.
 \end{abstract}


\maketitle

\emph{Introduction} - 
The response of materials to electromagnetic fields is
determined by electronic excitations that are strongly influenced
by electron-hole (e-h) interactions.
In particular, the e-h attraction leads to the formation of excitons, which are a fundamental aspect in the functionality of many optoelectronic devices, 
as excitons can propagate in materials carrying excitation energy that can be transformed and
exploited by different means. 
Excitons can be identified in  electronic spectra
as sharp peaks within the band gap  of insulators \cite{Knox1963,Bassani1975} or, beyond the band gap,
as spectral intensity enhancement towards lower energies
with respect to a non-interacting theoretical picture \cite{Hanke1979}.
A reliable description and analysis of those two-particle correlation effects
is therefore the key to understand materials' dielectric properties,
guide the realisation of new experiments and foster the development of new technological
applications. 

Nowadays, the {\it ab initio} solution of Bethe-Salpeter equation (BSE)  \cite{Strinati1988,Albrecht1998,Benedict1998,Rohlfing2000}
represents the state-of-the-art method 
to obtain spectra
in very good agreement with experiments 
in a large variety of materials \cite{Onida2002}. 
These theoretical achievements have mainly focused on 
optical absorption and electron energy-loss spectroscopy (EELS) spectra
for vanishing momentum transfer $\bfq\rightarrow0$.
However, judging whether a theoretical approach captures fully the physics of the electron dynamics requires
stringent tests that are offered by the measurement of the full dynamic range of the relevant variables (momentum and energy).
This assessment is now possible thanks to
the spectacular progress of inelastic x-ray scattering (IXS) experiments, both in the resonant (RIXS) and non-resonant (NRIXS) conditions. They 
allow one to probe electronic excitations at finite momenta ($\bfq$) \footnote{Atomic units are used throughout the paper.}
with a resolving power that has improved by orders of magnitude in the last two decades \cite{Schuelke2007,Rueff2010,Ament2011,Ishii2013}. 
The challenge for theory is hence the first-principles description of the full electron dynamics \cite{Schuelke2007}, 
which is well beyond the sole simulation of optical absorption spectra.
The investigation of the energy-momentum dispersion of elementary excitations (excitons, plasmons, etc.) provides fundamental information on
the way they propagate in materials.
Moreover, spectroscopic features measured at larger $\bfq$ 
allow for the study of excitations that in real space occur on shorter interatomic 
scales 
and/or are not visible in optics because they are dipole forbidden (see e.g. Refs. \cite{Larson2007,Verbeni2009,Hiraoka2011}).

Recently, some of us have shown that these ambitious goals 
are within reach, thanks to the extension of the BSE to describe e-h excitations carrying a finite momentum $\bfq$ \cite{Gatti}.
In the case of a prototypical wide-gap insulator such as lithium fluoride, 
the dynamic structure factor $S(\bfq,\w)$ was obtained in excellent agreement with accurate NRIXS data \cite{Abbamonte}.
However, in contrast to LiF, in hexagonal boron nitride (h-BN), a layered material that is
the insulating counterpart of graphite,
a recent comparison \cite{Galambosi} between BSE calculations and NRIXS data revealed a mismatch 
at high momentum transfers. The experiment displays a sharp peak at $\sim 7$ eV  
whose origin has still to be understood, as pointed out in Ref. \onlinecite{Galambosi}.
Moreover, despite the  accurate BSE results in LiF \cite{Gatti}, a recent study based on a simplified exciton kinetic kernel model \cite{Lee2013}
raised doubts about the capability of the {\it ab initio} BSE in general to address the issue of the exciton band structure.
Therefore new questions arise: Is LiF only a fortunate case for BSE? 
For  spectra at finite $\bfq$  should we hence go beyond the standard BSE implementation that has been successfully applied for optics?
Or, on the contrary, is the sharp peak measured in h-BN an artifact of the NRIXS experiment?
Finally, is  the BSE the appropriate method in general to study the exciton band structure of complex materials?

In the present work, we demonstrate that it 
is possible to reconcile theory and experiment
if the microscopic details of the screened Coulomb e-h interaction are explicitly taken into account in the calculations.
We show that in h-BN excitonic effects not only produce a redshift of the peaks, as observed in Ref. \onlinecite{Galambosi}, 
but also lead to important spectral shape redistributions.
We  reproduce all the details of the NRIXS spectra measured in Ref. \onlinecite{Galambosi}, including the 7 eV peak at large $\bfq$, 
and we consistently explain the appearance of excitonic features at various $\bfq$ 
as a multiple manifestation of the same large joint density of states (JDOS).
Moreover, by calculating the entire map of $S(\bfq,\w)$  in the first 3 Brillouin zones of h-BN, 
we identify new ``hot spots'' in dynamics, where the probability for creating an exciton is the largest. 
These predictions are fully confirmed by the new NRIXS experiment 
that we have performed at the beamline ID20 of the
European Synchrotron Radiation Facility (ESRF) in Grenoble (France).
Therefore these new results demonstrate that the combination of first-principles BSE calculations and accurate NRIXS experiments 
is a very powerful tool to explore and understand the exciton dynamics.


\emph{Theory} - The spectrum obtained in an NRIXS experiment is proportional to the dynamic structure factor: 
\beq
S(\bfq,\w) = -\frac{q^2}{(4\pi^2n)}\text{Im} \epsilon_M^{-1}(\bfq,\w),
\eeq
where $n$ is the average electron density. The
inverse of the macroscopic dielectric function $\epsilon_M^{-1}$  can be expressed as \cite{Gatti,Onida2002}:
\beq
\epsilon_M^{-1}(\bfq,\w) =  1 + \frac{8\pi}{q^2} \sum_\lambda \frac{\Big|\sum_{t} A_\lambda^{t}(\bfq)
\tilde{\rho}^t(\bfq)
 \Big|^2 }{\w-E_\lambda(\bfq) + i\eta},
\label{spectrumBSE}
\eeq
where in the Tamm-Dancoff approximation (TDA) \footnote{In solids effects beyond the TDA  generally are important for spectra involving  $\text{Re} \, \epsilon_M$ \protect\cite{Olevano2001}
and are neglected for $\text{Im} \, \epsilon_M$  \protect\cite{Onida2002}.}
the sum is over  valence-conduction ($v$-$c$) transitions $t$, the oscillator strengths are
$\tilde{\rho}^t(\bfq)=\bra{\phi_{v\bfk-\bfq_\bfr}}e^{-i\mathbf{q}\mathbf{r}}\ket{\phi_{c\bfk}}$,  $\bfk$ and $\bfq_\bfr$ are in the first Brillouin zone, 
and $\bfq=\bfq_\bfr+\bfG$ is the measured momentum transfer with  a reciprocal-lattice vector $\bfG$.
The BSE can be cast into  an effective two-particle Schr\"odinger equation  \cite{Gatti,Onida2002}:
$H_{\textrm{exc}}(\bfq) A_\lambda(\bfq) = E_\lambda(\bfq) A_\lambda(\bfq)$, where $A_\lambda(\bfq)$ and $E_\lambda(\bfq)$ are the exciton eigenvectors and eigenvalues, respectively.
The exciton hamiltonian $H_{\textrm{exc}}$:
$\bra{t} H_{\textrm{exc}}\ket{t'}=  
E_t \delta_{t,t'} + \bra{t} v_c-W \ket{t'}$
contains  the quasiparticle e-h transition energies $E_{t}$ calculated in the GW approximation \cite{Hedin1965}, and the matrix elements in the transition basis of 
the bare Coulomb interaction $v_c$  and  the statically screened Coulomb interaction $W=\epsilon^{-1}v_c$, which describes the e-h attraction (i.e. gives rise to excitons) and is here obtained in the random phase approximation (RPA). 
We have calculated the ground state of h-BN within the local density approximation (LDA) \protect\cite{KS} of density functional theory \protect\cite{HK}, 
using norm-conserving Troulliers-Martin pseudopotentials \protect\cite{TrM} in a plane wave approach \protect\cite{Abinit} with an energy cutoff of 30 Hartree.
In order to approximate  GW quasiparticle energies, following \protect\cite{Wirtz} we corrected the LDA band structure with a scissor operator of 1.98 eV for the band gap and  stretching the valence bands by 5\% .
The BSE spectra at finite $\bfq$ are obtained from the EXC code \protect\cite{EXC} using 25 bands and a $18\times18\times4$ $\bfk$-point grid for calculations with $\bfq$ parallel to the hexagonal plane and with a  $12\times12\times8$ $\bfk$-point grid for calculations with $\bfq$ perpendicular to it.

\emph{Experiment} - In the NRIXS experiment the incident x-ray beam from three undulators was monochromated to an
energy of 7.5 keV by a combination of a Si(111) double crystal and a Si(311) channel-cut. The
beam was focused to a spot of $\approx$ 10 $\mu$m $\times$ 20 $\mu$m
(V $\times$ H).
The spectrometer used a diced Si(533) analyzer crystal in the Johann
geometry with a Rowland circle diameter of 2 m. The active diameter
of the analyzer crystal was 80 mm, yielding a momentum-transfer
resolution of $\sim$ 0.15 \AA$^{-1}$. The energy resolution was 200 
meV (FWHM). The scattering plane was
vertical, i.e., perpendicular to the plane of linear polarization of
the incident and scattered radiation. The detector was based on a
Timepix chip, with a pixel size of 55 $\mu$m and enabling us to use
the dispersion compensation algorithm \cite{Huotari2005,Huotari2006}.
The analyzer Bragg angle was fixed at 87$^\circ$ and the energy-loss
spectra were measured by scanning the incident-photon energy.
The h-BN single crystal was a colorless and transparent platelet with a thickness of 0.5 mm and lateral size of 2.1 mm, which was produced by a high- pressure and high-temperature (HPHT) method using a barium-related solvent system as reported in \cite{Watanabe2}.
The samples were aligned using x-ray diffraction on the beamline.

\emph{Results} - In layered h-BN the electronic states can be classified according to the even ($\sigma$) and odd ($\pi$) parity with respect to the single BN sheet \cite{Doni1969}.
 Optical and EELS spectra \cite{Buechner1977,Mamy1981,Tarrio1989,Watanabe1} in the low-energy range, 
where excitonic effects are more relevant, are determined  by  $\pi-\pi^*$ transitions that are mainly visible for in-plane $\bfq\rightarrow0$. 
The weak screening of the Coulomb e-h interaction, in concomitance with a JDOS peak
  due to vertical transitions between parallel bands \cite{supmat} in the HKML plane
 of the Brillouin zone \cite{Doni1969,Catellani1985},
 in the $\bfq\rightarrow0$  spectra  gives rise  to prominent exciton peaks with large binding energies \cite{Arnaud,Wirtz,Wirtz2005}.
 
\begin{figure}
\includegraphics[width= 0.49\textwidth, angle=0]{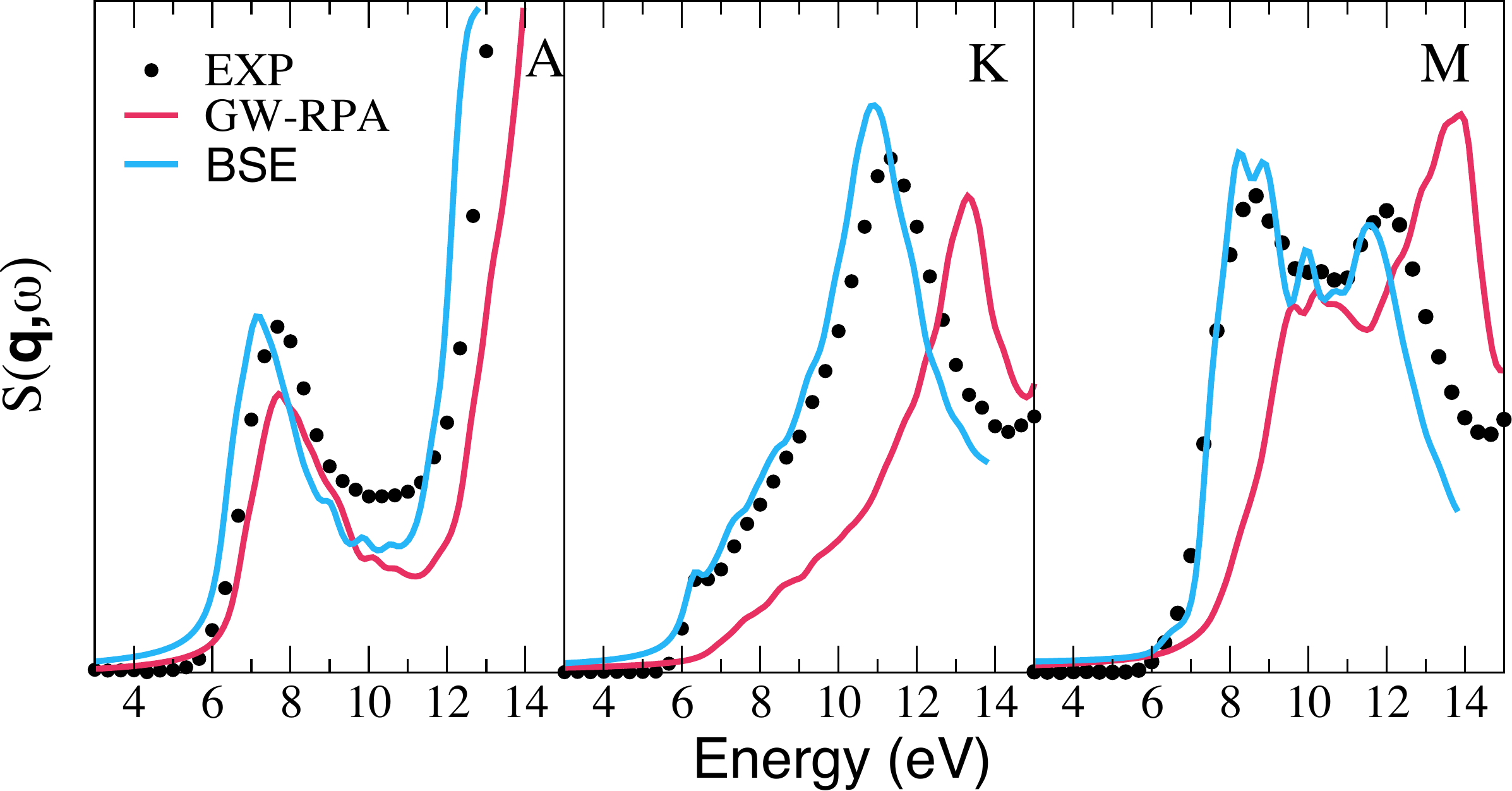}
\begin{center}
\caption{Dynamic structure factor at three Brillouin zone boundaries $\bfq=$ A, K, and M  
calculated in the GW-RPA 
and from the solution of the BSE 
compared to the NRIXS data 
from Ref. \protect\cite{Galambosi}.}
\label{fig1}
\end{center}
\end{figure}

Fig. \ref{fig1} shows the dynamic structure factor $S(\bfq,\w)$ in the energy range of the $\pi-\pi^*$ transitions 
at three finite momentum transfers: $\bfq=$  A, K, and M, which are located  at the boundaries of the first Brillouin zone \cite{supmat}. 
As noticed in \cite{Galambosi}, the e-h attraction induces a redshift of the peaks, which is larger for the in-plane $\bfq$ directions at K and M (redshift of 1.5 eV) than for $\bfq$ at A  (redshift of 0.4 eV), i.e. along the direction perpendicular to the BN layers,
implying an anisotropic effect of the e-h interaction 
\footnote{It is also interesting to note that this anisotropy does not correlate directly with the dielectric constant, 
which is smaller for $\bfq$ perpendicular to the layers than for in-plane $\bfq$ (2.53  vs. 4.40 in RPA \cite{Arnaud}).
Although one would hence expect asmaller e-h attraction at $\bfq=$ K and M, the resulting larger binding energies  are instead due to a larger spatial confinement of the e-h pairs.}.
This is evident from the comparison of the BSE spectra  
with the GW-RPA results, 
obtained starting from the GW band structure and neglecting 
the e-h attraction $W$ in the exciton hamiltonian $H_{\textrm{exc}}$.
At variance with \cite{Galambosi}, where the shift was inferred from the adjustment of the calculations to the experimental spectra,
the anisotropic redshift of the spectra is here the direct outcome of the BSE calculations that result in very good agreement with NRIXS. 
Our simulations of the experimental spectra, being free from any adjustable parameters, allow us to additionally observe that 
excitonic effects also induce an important redistribution of the spectral weight towards lower energies with respect to the non-interacting e-h picture.

At $\bfq=$ M  the two main peaks are located  at $\sim$ 8 and $\sim$ 12 eV.
At the K point, the first peak at $\sim$ 6-7 eV appears  as a shoulder of the second one.
In both cases the 12 eV peak originates from non-vertical (i.e. $\bfk \rightarrow \bfk'=\bfk+\bfq$) $\pi-\pi^*$ transitions that disperse isotropically as a function of in-plane $\bfq$, 
shifting the energy of the peak from  9 eV at $\bfq\rightarrow0$ to 12 eV in both $\Gamma$K and $\Gamma$M  directions.
The first peak, instead, is due to a peculiar property of the hexagonal Brillouin zone,  see Fig. \ref{fig4}(a).
For $\bfq=$M it derives from \emph{vertical} e-h transitions between $\bfk$ points belonging to the ML line in the band structure \cite{supmat} 
(analogous, for $\bfq$=K, is the HK line).
These are the same vertical $\pi-\pi^*$ transitions with large JDOS  that are at the origin of the tightly bound exciton in the spectra at $\bfq\rightarrow0$ \cite{Arnaud,Wirtz}.
The same strong excitonic effects are hence appearing also at $\bfq=$M and K as a spectral shape redistribution that similarly strongly enhances the first peak in the spectrum. They create a new shoulder at $\bfq=$K (entirely absent in GW-RPA) and  at $\bfq=$M they make the first peak become more intense than the second one, in contrast to the GW-RPA results.

\begin{figure}
\begin{center}
\includegraphics[width= 0.3\textwidth, angle=0]{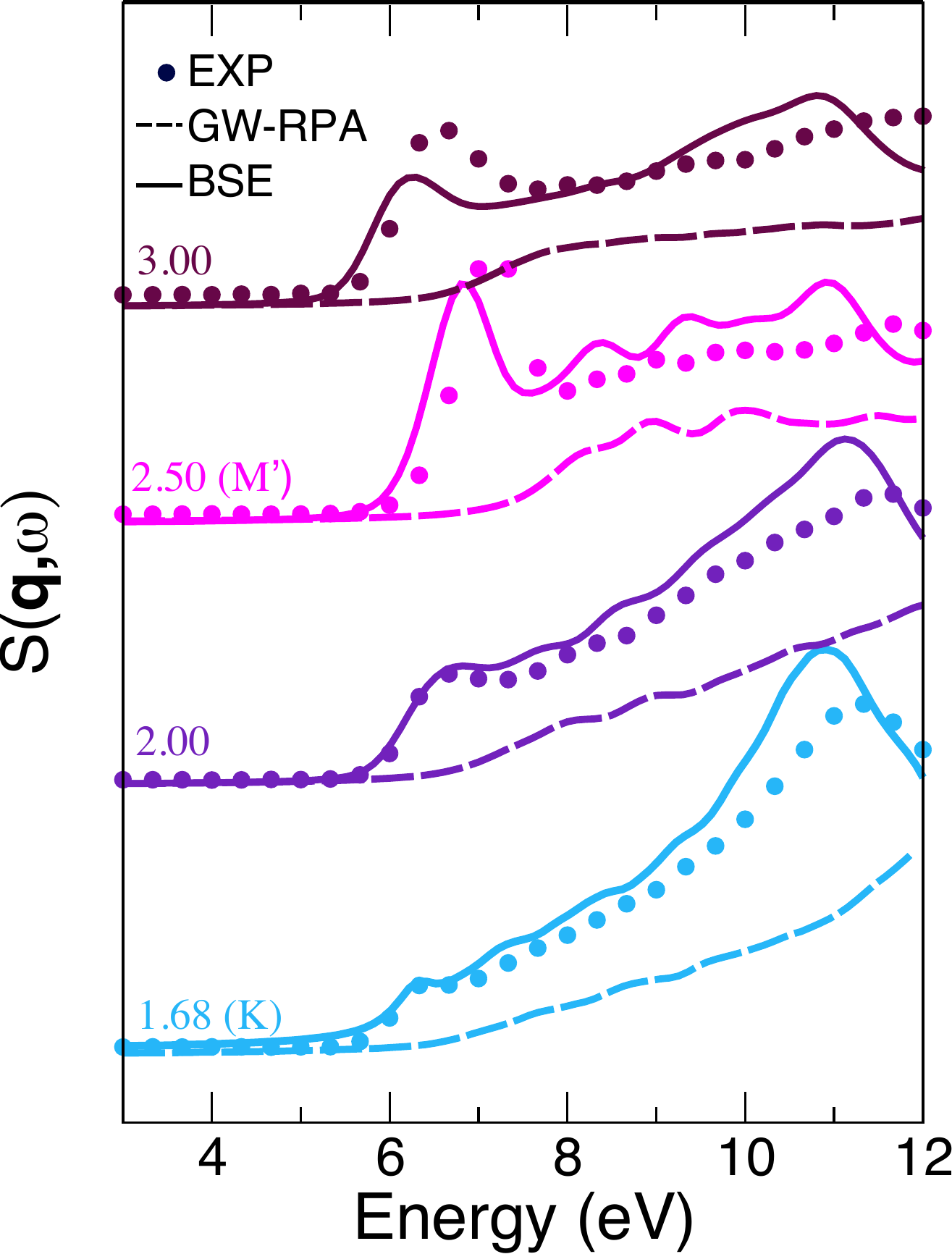}
\caption{Dynamic structure factor for $\bfq$ along  $\Gamma$K direction calculated in the GW-RPA 
and from the solution of the BSE 
compared to the NRIXS data 
from Ref. \protect\cite{Galambosi}. 
The $\bfq =$ 2.5 \AA$^{-1}$ is the M$'$ point (see text). }
\label{fig2}
\end{center}
\end{figure}

\begin{figure*}
\begin{center}
\includegraphics[width= 0.9\textwidth, angle=0]{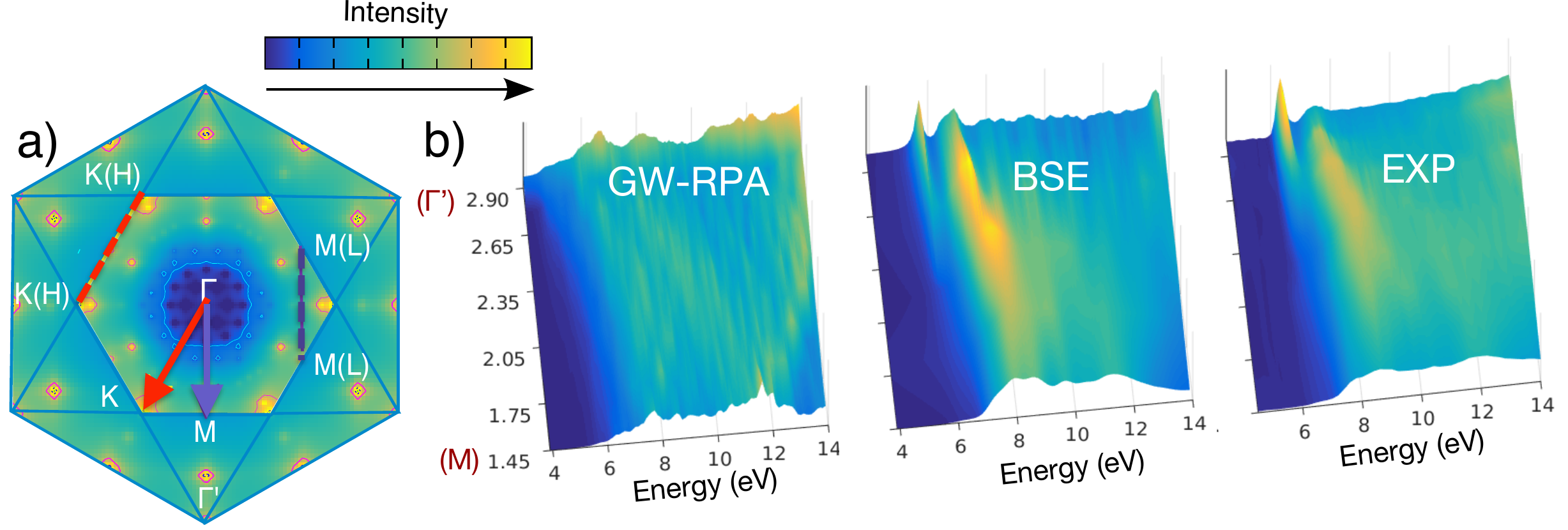}
\caption{(a) Color map of $S(\bfq,\w)$ calculated from the BSE for fixed $\w=$ 7 eV and for all $\bfq$ spanning the first 3  Brillouin zones with $\bfq_z=0$. To enhance its visibility, $S(\bfq,\w)$ in the first BZ has been magnified by a factor 3.
The high-symmetry points relevant for the discussion in the main text are explicitly indicated (in parenthesis those belonging to the $\bfq_z=0.5$ plane). 
Adding $\bfq=\Gamma$M (solid violet line) to $\bfk=$M (L) one obtains another $\bfk'=\bfk+\bfq =$M (L) (see dashed violet line). Equivalently adding $\bfq=\Gamma$K  (solid red line) to $\bfk=$K (H) one obtains again $\bfk'=$K (H)  (see red dashed line). 
(b) Comparison between calculated and measured $S(\bfq,\w)$ for $\bfq$ along the $\Gamma$M direction.}
\label{fig4}
\end{center}
\end{figure*}

Having clarified the diverse role of excitonic effects at various $\bfq$, we are ready to discuss what happens along the $\Gamma$K direction, 
where for $\bfq > 2.0$ \AA$^{-1}$ a peak appears in the NRIXS spectra at $\sim$ 7 eV that was unexplained in Ref.  \cite{Galambosi}. 
The spectra  
that we have calculated for different momentum transfers along $\Gamma$K  demonstrate that the BSE is actually able to reproduce the 7 eV peak that is measured in the NRIXS experiment  \cite{Galambosi} (see Fig. \ref{fig2}). 
Moreover they show that the peak has indeed an excitonic character as it is completely absent in the GW-RPA calculations. 
In order to better understand the origin of this excitonic peak one has to note that it has the largest intensity at $\bfq =$ 2.5 \AA$^{-1}$.
This momentum transfer along the $\Gamma$K direction is in fact another M point, which is located at the boundary of the third Brillouin zone and which we call M$'$.
We can also immediately recognise why excitonic effects are again particularly strong at this momentum transfer M$'$. 
The exciton eigenvalues $E_\lambda(\bfq)$ and eigenvectors $A_\lambda(\bfq)$ at M$'$ must be the same as at the M point in the first Brillouin zone 
since the two points just differ by a reciprocal-lattice vector $\bfG$.
However the oscillator strengths $\tilde{\rho}^n(\bfq)$ are generally different at different $\bfq$ and this explains why the spectra at M and M$'$ do not entirely overlap. 
In particular, at $\bfq$=M$'$ there is a prominent peak at 7 eV emerging from a featureless plateau at higher energy, while at $\bfq$=M a double-peak structure was observed in the same energy range.

We have therefore found that the solution of the BSE, when the screening of the Coulomb interaction is explicitly calculated at the RPA level rather than obtained from model dielectric functions \cite{Galambosi,Soininen2000,modelW,modelW2}, 
is able to reproduce and explain the NRIXS data.
This is confirmed by the direct comparison between the calculated and experimental spectra \cite{supmat}. The agreement is excellent in all considered cases.
Moreover, having understood that strong excitonic effects for the peak at 7 eV should be expected whenever $\pi-\pi^*$ transitions give rise to an intense JDOS,
we can forecast other ``hot spots'' with large intensity in the dynamic structure factor, beyond what has been experimentally detected in  \cite{Galambosi}.
In Fig. \ref{fig4}(a) we thus plot a color map of $S(\bfq,\w)$ for a fixed $\w =7$ eV and for all $\bfq$ spanning the first 3 $\bfq_z=0$ Brillouin zones.  In this manner we can easily identify  two other remarkable points along the $\Gamma$M direction, located between $\bfq$=M and the vertex of the hexagon $\bfq$=$\Gamma'$, which differs from the $\Gamma$ point by a reciprocal-lattice vector $\bfG$.

In order to confirm these predictions and assess the whole theoretical map of $S(\bfq,\w)$ that has been obtained from the BSE calculations,
we compare the theoretical results to those obtained from our NRIXS experiment in 
Fig. \ref{fig4}(b). While the GW-RPA results are totally different from the experiment, 
the excellent agreement between the BSE results and the NRIXS data proves the predictive power of the BSE.
 In the spectra  the largest intensity is found at $\bfq=\Gamma'$, 
displaying two prominent peaks  deriving again from the large JDOS of vertical $\pi-\pi^*$ transitions.
The first peak, visible only close to $\bfq=\Gamma'$, matches the main exciton peak at the onset of the spectrum at $\bfq\rightarrow0$.
The second peak at $\bfq=\Gamma'$, which is not noticeable at $\bfq\rightarrow0$, instead evolves continuously from the first peak located at $\sim$ 8 eV at $\bfq$=M.
These results show that inspecting spectra at large $\bfq$, thanks to the variation of the oscillator strengths with $\bfq$, 
can reveal formation of excitons that are hidden in corresponding optical spectra.
Moreover, they evidence how the combination of NRIXS and BSE provides the means for their detection in the whole energy-momentum range, 
allowing one to obtain the full exciton band structure.

\emph{Conclusions} - In summary, by solving the discrepancies between previous experiments and calculations,
we have successfully established  the energy-momentum map covering the first 3 Brillouin zones
of the dynamic structure factor $S(\bfq,\w)$ of h-BN, a prototypical layered insulator. 
We have shown that excitonic effects in h-BN are strong also at large momentum $\bfq$. 
They are essential to interpret and understand the spectra (including previous unexplained features) 
and cannot be neglected remaining at a level of theory corresponding to the RPA \cite{Galambosi}.
We have explicitly proved that such a theoretical map based on BSE is crucial in order to guide the experimental exploration of the electronic dynamics, as new IXS measurements have fully validated the theoretical predictions.
Therefore these case-study results  promote the {\it ab initio} solution of the BSE, which has been already successfully applied in optics, as an accurate and predictive method also to investigate the charge dynamics and the exciton band structure for a wide range of materials and technological applications.

This research was supported by a Marie Curie FP7 Integration Grant within the 7th European Union Framework Programme and by Academy of Finland (contract numbers 1259599, 1260204, 1254065, 1283136, 1259526).
Computational time was granted by GENCI (Project No. 544).

\bibliographystyle{apsrev4-1}
\bibliography{biblio}

\end{document}